\documentclass[preprint,showpacs,preprintnumbers,amsmath,amssymb,10pt,aps,pra]{revtex4}

\usepackage{graphicx}
\usepackage{amsmath}
\usepackage{color}

\begin{document}

\title{Atom interferometry using $\delta$-kicked and finite duration pulse-sequences}
	\author{Boris Daszuta}
\author{Mikkel F. Andersen}
		\affiliation{Jack Dodd Centre for Quantum Technology, Department of Physics, University of Otago, New Zealand}	
	\begin{abstract}

We investigate an atom interferometer in which large momentum differences between the arms are obtained by using quantum resonances in the atom optics $\delta$-kicked rotor. The interferometer can potentially measure the Talbot time (from which $h/m$ can be deduced), the local gravitational field, or can serve as a narrow velocity filter. We present an analytical analysis in the short pulse limit, and a numerical investigation for finite pulse durations. The sensitivity of the interferometer is improved by a moderate violation of the short pulse limit. Remarkably simple relations predict the optimal pulse duration, and the sensitivity of the interferometer.
	\end{abstract}
	
\date{Month Day, Year}

\maketitle

\section{Introduction}
Atom interferometers provide an exciting tool for precision measurements \cite{berman1996atom, cronin2009optics}. In recent years applications range from a new determination of the fine structure constant $\alpha$ \cite{cadoret2008combination} and measurements of Newton's gravitational constant $G$ \cite{fixler2007atom} to measurements of the local gravitational field at both micro-meter \cite{Poli2011} and larger length scales \cite{Merlet2010}.

In light-pulse atom interferometers matter waves are split and reflected using momentum transfer between light and atoms \cite{berman1996atom, cronin2009optics}. The basis for such an atom interferometer is the coherent splitting of a cold atomic wavepacket into two or more arms, followed by phase accumulation and subsequent recombination of the arms. In these schemes the interferometer measures the difference in phase between the matter waves in the different arms. The phase accumulated by an ensemble of atoms with mass $m$ and momentum eigenvalue $p$ in a time interval $T$ is given by $\frac{p^2}{2m\hbar}T$. When increasing the momentum difference $\Delta p$ between arms, the sensitivity of an atom interferometer can therefore be improved as the square of $\Delta p$.

Significant effort is expended at investigating atom interferometers where the arms are separated by a large $\Delta p$ \cite{Gupta2002}. Interferometers that use high order diffraction processes \cite{muller2008atom, sfcharexperTaluk2010} suffer a technical limitation as the required laser power increases sharply with $\Delta p$ \cite{mullertheory2008}. This problem has been overcome using Bloch oscillations to accelerate the atoms in the arms of the interferometer \cite{clade2009large, mullerembedded2009}. Bloch oscillations are an adiabatic process and this scheme therefore does not suffer the laser power limitations of high order diffraction processes. However, being an adiabatic process the interaction time between the light and the atoms is significant which can lead to drifts or noise that limits the accuracy of measurements performed \cite{clade2009large}. To suppress this a symmetric configuration has been demonstrated \cite{mullerembedded2009}. An alternative scheme is to use a sequence of Bragg pulses where \cite{Chiow2011} reports on an interferometer with a momentum difference between arms of $102$ photon recoil momenta.

In this paper we investigate an atom interferometer that builds on the principles of the one considered in \cite{sfcharexperTaluk2010, mcdowall2009fidelity}. However, we omit the need for a high order diffraction pulse by replacing it with a train of low order pulses. In analogy to the sequential Bragg interferometer \cite{Chiow2011}, a large $\Delta p$ is obtained without utilizing a high order diffraction pulse or an adiabatic process, but through series of low order diffraction pulses.
The operating principle of the interferometer relies on quantum resonances in the atom optics $\delta$-kicked rotor, which are briefly described in section \ref{sec:QR}. In order to obtain an in-depth understanding of the proposed interferometer section \ref{sec:modelSys} introduces an idealized model based on diffraction pulses in the ``short pulse'' or ``Raman-Nath'' limit. In sections \ref{qdkr} and \ref{qdka} we analytically investigate the performance of the idealized interferometer for measuring the ``Talbot time'' (which together with other well known constants constitutes a measurement of the fine structure constant), the initial momentum of the input matter wave, and the local gravitational field. In section \ref{sec:longpulse} we numerically investigate how the interferometer performs when the Raman-Nath condition is violated. We find that violating the Raman-Nath condition improves the sensitivity of the interferometer, with remarkably simple relations valid over a large range of parameters predicting the optimal pulse duration as a function of other parameters.

\section{Quantum resonances in the atom-optics $\delta$-kicked rotor}\label{sec:QR}
The atom-optics $\delta$-kicked rotor \cite{Moore1994, Moore1995} is a kicked particle realized using laser cooled atoms subjected to periodic pulses of a standing wave of laser light. The laser light is assumed to be sufficiently far off resonance such that spontaneous photon scattering can be ignored. Through the optical dipole force the standing wave forms a spatially periodic potential with a period of half the wavelength of the light used for its generation. If the pulses of the optical potential are sufficiently short, such that the distance an atom travels during the pulse is much smaller than the period of the potential---the so-called Raman-Nath regime, then the 1D dynamics of the atoms are well described by the idealized atom-optics kicked rotor Hamiltonian given by \cite{Raizen1999}:
\begin{equation}
 \hat{H}\left(t\right)= \frac{\hat{p}^2}{2 m}+\hbar \phi_d \cos \left(\kappa \hat{x} \right)\sum_{n=0}^{N-1}\delta \left( t-nT \right),
\label{eq:krh}
\end{equation}
where $\hat{x}$ and $\hat{p}$ are the canonical position and momentum operators respectively, $t$ the time, $T$ the time between pulses, $m$ the mass of the atom, $\kappa=2k_L$ is twice the wave number of the light ($k_L$) forming the potential, $\phi_d$ is the kicking strength, and $N$ the number of pulses. The stroboscopic time evolution of an initial state due to Eq. \eqref{eq:krh}, may be described by repeated applications of the kick-to-kick evolution operator:
\begin{equation}\label{eq:tevopkick}
\hat{U}=\exp\left(-\frac{i}{\hbar}\frac{\hat{p}^{2}}{2m}T\right)\exp\left(-i\phi_{d}\cos\left(\kappa \hat{x}\right)\right).
\end{equation}
Eq. \eqref{eq:tevopkick} factorizes as a product of a kick operator and a free space evolution operator due to the $\delta$-function time dependence in the Hamiltonian (Eq. \eqref{eq:krh}).

Quantum resonances in the $\delta$-kicked rotor occur due to the matter wave Talbot effect \cite{temporalTalbot} for specific combinations of initial atomic momentum and the time between pulses $T$. We first consider the case where the atom is initially in a zero momentum eigenstate ($ \left| 0 \right\rangle $) and the ``first quantum resonance'', for which $T = T_T = 2 \pi /(4\omega_r)$, where $T_T$ is the Talbot time, and $\omega_r = \hbar k_L^2 /(2m)$ is the atomic recoil frequency. The resonance occurs as follows: The kick operator imprints a spatially periodic phase on the initial wave function, thereby imparting momentum to the atoms in integer multiples of $\hbar \kappa$ by diffraction. During the free evolution between kicks each such momentum component $q\hbar \kappa$ ($q \in \mathbb{Z}$), will acquire a phase given by $\Phi_q=\frac{q^2 \hbar \kappa^2}{2m}T_T=q^2 2 \pi$. Since $q$ is an integer, all phases are integer multiples of $2 \pi$ and a revival of the wavefunction occurs \cite{temporalTalbot}. The free-space evolution term in Eq. \eqref{eq:tevopkick} therefore reduces to the identity operator, hence we conclude that the kicks add coherently and consecutive pulses serve to increase the amplitude of the phase modulation. This leads to ballistic transfer of momentum to the atoms and the wave function spreads linearly in momentum space with $N$ \cite{qres2003, izrailev1980quantum, Ryu2006}.

The $\delta$-kicked rotor can be used for making sub-Fourier measurements \cite{sfcharexperTaluk2010, mcdowall2009fidelity, qres2003, Ryu2006, szriftgiser2002observation, Tonyushkin2008, Horne2011}. In \cite{sfcharexperTaluk2010, mcdowall2009fidelity} it was shown that identifying the Talbot time by measuring the kinetic energy or momentum imparted to the atoms around the quantum resonance is not the most sensitive measurement. A superior alternative is to add a high order diffraction pulse at the end of the sequence, that ensures interference between all populated momentum states \cite{sfcharexperTaluk2010}.

\section{Interferometer Description}\label{sec:modelSys}
Our present goal is to identify an atom interferometer that use quantum resonances in the atom optics $\delta$-kicked rotor as a ``beam splitter'' to split a wave packet into momentum components that differ largely. At the same time it should not rely on a high order diffraction pulse as in \cite{sfcharexperTaluk2010, mcdowall2009fidelity} because laser power then poses a technical limit on the momentum difference that can be obtained. 

\begin{figure}
\includegraphics[width=8.6cm]{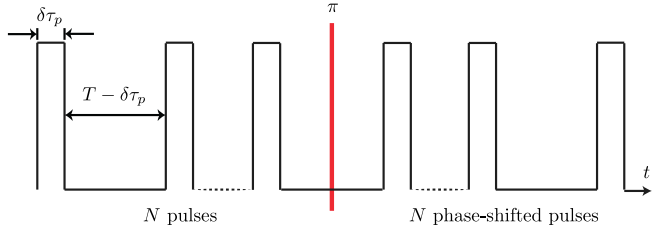} 
\caption{(Color online) ``Anti-symmetrized'' pulse train schematic: Each pulse is applied for a duration $\delta\tau_p$ (which in the Raman-Nath regime corresponds to a $\delta$-kick of strength $\phi_d$) to an initial state $\left|\psi_i,t=0\right>$. The state subsequently undergos freespace evolution for $T-\delta \tau_{p}$. A total of $N$ pulse-freespace evolutions are performed. After the first pulse train (the end of which is denoted by the solid vertical line) the optical potential is $\pi$-phase shifted and a secondary train of $N$ pulses (kicks) is applied, yielding the final (output) state $\left|\psi_f,t=2NT\right>$.} 
\label{fig:pulseTrainSchematic}
\end{figure}

To achieve this we use two consecutive kicked rotor pulse sequences where the optical standing wave has had a $\pi$ phase shift between the two sequences (see Fig. \ref{fig:pulseTrainSchematic}). Such a capability has been demonstrated experimentally \cite{sfcharexperTaluk2010, Ullah2011, Sadgrove2011}. We ignore interactions between atoms, which is valid either when using a dilute sample or a Feshbach resonance to make the scattering length negligible \cite{inouye1998observation, feshbachrev}. The Hamiltonian governing the evolution of a state exposed to the proposed ``anti-symmetrized'' kicked rotor sequence (ASKRS) is then given by:
\begin{equation}\label{eq:hamdeltakick}
\hat{H}_{\delta k}(t)=\frac{\hat{p}^{2}}{2m}+\hbar\phi_{d}\cos\left(\kappa\hat{x}\right)\left(\sum_{n=0}^{N-1}\delta\left(t-nT\right)-\sum_{n=N}^{2N-1}\delta(t-nT) \right).
\end{equation}
The stroboscopic time evolution of an initial state may be described by $N$ repeated applications of the kick-to-kick operator of Eq. \eqref{eq:tevopkick}, followed by $N$ applications of its ``$\pi$-phase shifted'' counterpart:
\begin{equation}\label{eq:kickopv}
\hat{V}=\exp\left(-\frac{i}{\hbar}\frac{\hat{p}^{2}}{2m}T\right)\exp\left(i\phi_{d}\cos\left(\kappa\hat{x}\right)\right).
\end{equation}
For an initial state $\left|\psi_i,t=0\right>$ the final state after evolution in the ASKRS of Eq. \eqref{eq:hamdeltakick} $\left|\psi_f,t=2NT\right>$ will be given by:
\begin{equation}
\left|\psi_f,t=2NT\right>=\hat{V}^N\hat{U}^N\left|\psi_i,t=0\right> .
\end{equation}
The output of the interferometer is a measurement of the fraction of atoms that has had their momentum returned to their initial momentum after the two pulse sequences. For an atom initially in a momentum eigenstate $\left|p_0\right>$ the interferometer output is therefore: 
\begin{equation}\label{eq:interoutput}
I\left(\varepsilon , p_0 \right)=\left|\left<p_0\right|\hat{V}^N\hat{U}^N\left|p_0\right>\right|^2,
\end{equation}
where $\varepsilon=T-T_T$. In order to measure $I$ the momentum spread of the initial sample must be less than the photon recoil momentum of the light used to form the standing wave potential. This can be achieved either by using a Bose-Einstein condensate \cite{Ryu2006, sfcharexperTaluk2010} or a velocity selected thermal sample \cite{cadoret2008combination, muller2008atom}.  

 \par
\begin{figure}
\includegraphics[width=8.6cm]{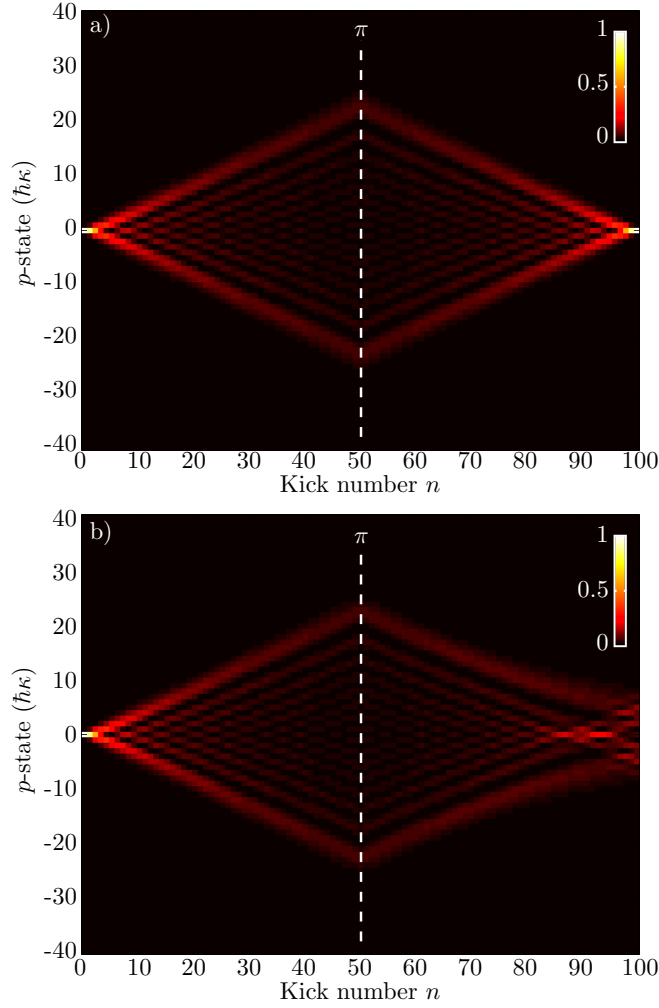} 
\caption{(Color online) (a) Coherently imparted momentum to an initial state $\left|0\right>$ results in a linear spread of occupied momentum state populations, with each state being separated by an integer multiple of $\hbar \kappa$. The population of states with ever higher magnitude of momentum (and hence energy) grows during application of the first pulse train. Subsequently the $\pi$-shifted train is applied (dashed vertical line), resonant kicking allows for coherent recombination, i.e. a perfect echo. (b) Deviation from $T_T$ by $\varepsilon=3\mathrm{ns}$, yields relative phases that obstruct complete recombination. Parameters: (a) $\lambda=780\mathrm{nm}$, ${}^{85}M_\mathrm{Rb}$, $\phi_d=0.5$, $\varepsilon=0$, $N=50$ (b) $\varepsilon=3ns$.} 
\label{fig:pPopViz}
\end{figure}

 Figure \ref{fig:pPopViz} illustrates the operation of the pulse sequence in momentum-space. It shows the probability density in color code on the discrete set of momenta that is coupled, as a function of kick number. During the first sequence the evolution of an initial state is described in section \ref{sec:QR}. For exact resonance ($\left|\psi_i, t=0\right>=\left|0\right>$ and $T=T_T$) shown in Fig. \ref{fig:pPopViz}(a) the wavefunction right after a pulse is phase-modulated with an amplitude that is linearly increasing with pulse number, resulting in the desired splitting between the multiple ``arms'' of the interferometer. After the first $N$ pulses the phase of the diffracting standing wave potential is shifted by $\pi$. As the phase imprint will now have opposite sign, each pulse serves to reduce the phase modulation of the wavefunction. Thus at $2N$ the wavefunction will return to the initial momentum eigenstate. Under these conditions the $\pi$ phase shift of the standing wave potential works as an effective time reversal and the final state is a perfect ``echo'' of the initial. The ASKRS can be interpreted as a multi-path interferometer where the kicks couple different populated momentum states, which then acquire phase in between pulses. The echo described above therefore crucially relies on the phase accumulated between pulses being an integer multiple of $2 \pi$ for all populated momentum states. For a slight deviation of $\varepsilon$ in $T$ from $T_T$ (or of $p_0$ from $0$) the echo behavior is spoiled, resulting in a reduced return probability as shown in Fig. \ref{fig:pPopViz}(b). 

\section{Deviation From Resonance in $T$ and $p_0$}\label{qdkr}
We will now quantitatively consider the performance of the proposed interferometer (governed by Eq. \eqref{eq:hamdeltakick}) operating in the vicinity of the quantum resonance. To this end, we are interested in the effects of an initial state with a ``small'' non-zero momentum eigenvalue of $p_0$ and non-zero $\varepsilon$. 
\par Equation \eqref{eq:interoutput} yields the interferometer output $I\left(\varepsilon,p_{0}\right)$ under the aforementioned conditions. The dependence on $\varepsilon$ is only introduced through the kick operators $\hat{U}(\varepsilon)$ and $\hat{V}(\varepsilon)$. 
Due to the spatial periodicity of the Hamiltonian in Eq. \eqref{eq:hamdeltakick}, Bloch's theorem ensures that only a discrete set of momentum eigenstates of the form $p=q\hbar\kappa+p_0$ ($q\in\mathbb{Z}$) are coupled by the kick operators \cite{ashcroftss1976}; this yields the closure relation $\sum_{q\in\mathbb{Z}} \left|p_0+q\hbar\kappa\right\rangle \left\langle p_0+q\hbar\kappa\right | = \hat{\mathbb{I}}$ for each choice of $p_0$. 
In order to evaluate Eq. \eqref{eq:interoutput} we insert the closure relation and perform the calculation in two parts:
Explicitly, the state generated after $N$ applications of $\hat{U}\left(\varepsilon\right)$ to $\left|p_0\right>$ may be written as $ \hat{U}^N\left(\varepsilon\right)\left|p_0\right>=\sum_{q\in\mathbb{Z}} c_q\left(\varepsilon,p_0\right)\left|q\hbar\kappa+p_0\right>$, where the expansion coefficients $c_q\left(\varepsilon,p_0\right)$ are:
\begin{equation}\label{eq:cncoeff}
c_q\left(\varepsilon,p_0\right)=\left<p_0+q\hbar \kappa\right| \hat{U}^N(\varepsilon)\left|p_0\right>,
\end{equation}
Analogously for the second pulse train we define:
\begin{equation}\label{eq:dncoeff}
d_q^*\left(\varepsilon,p_0\right)=\left<p_0+q\hbar \kappa\right| \left( \hat{V}^\dagger(\varepsilon)\right)^N\left|p_0\right>^*.
\end{equation}
Equations \eqref{eq:cncoeff} and \eqref{eq:dncoeff} allow Eq. \eqref{eq:interoutput} to be rewritten in the form:
\begin{equation}\label{eq:Ioutcoeff}
I(\varepsilon,p_0)=\left|\sum_{q\in\mathbb{Z}}d_{q}^*(\varepsilon,p_0) c_{q}(\varepsilon,p_0)\right|^{2},
\end{equation} 
In order to evaluate this expression, we write the expansion coefficients in polar form as:
\begin{eqnarray}\label{eq:cncplxrot}
c_q\left(\varepsilon,p_0\right) & =  & A_q\left(\varepsilon,p_0\right)\exp\left(i\theta_q\left(\varepsilon,p_0\right)\right),\\\label{eq:dncplxrot}
d_q^*\left(\varepsilon,p_0\right) & =  & B_q\left(\varepsilon,p_0\right)\exp\left(-i\chi_q\left(\varepsilon,p_0\right)\right),
\end{eqnarray}
where $A_q\left(\varepsilon,p_0\right)$, $B_q\left(\varepsilon,p_0\right)$, $\theta_q\left(\varepsilon,p_0\right)$ and $\chi_q\left(\varepsilon,p_0\right)$ are real-valued functions.
For parameters close to resonance a first order Taylor expansion in $\varepsilon$ and $p_0$ accurately determines the aforementioned functions. Following \cite{mcdowall2009fidelity}, this leads to:
\begin{eqnarray}\label{eq:cncplxrot2}
c_q\left(\varepsilon,p_0\right) & \simeq & \mathrm{J}_q\left(N\phi_d\right)\exp\left(i\left(\left.\frac{\partial\theta_q}{\partial\varepsilon}\right|_{\varepsilon=0,\,p_0=0}\varepsilon+\left.\frac{\partial\theta_q}{\partial p_0}\right|_{\varepsilon=0,\,p_0=0}p_0-q\frac{\pi}{2}\right)\right),\\\label{eq:dncplxrot2}
d_q^*\left(\varepsilon,p_0\right) & \simeq & \mathrm{J}_q\left(N\phi_d\right)\exp\left(-i\left(\left.\frac{\partial\chi_q}{\partial\varepsilon}\right|_{\varepsilon=0,\,p_0=0}\varepsilon+\left.\frac{\partial\chi_q}{\partial p_0}\right|_{\varepsilon=0,\,p_0=0}p_0-q\frac{\pi}{2}\right)\right), 
\end{eqnarray}
where $\mathrm{J}_q$ is a Bessel function of the first kind of order $q$. 
In particular, the functional dependence of Eq. \eqref{eq:cncplxrot2} and \eqref{eq:dncplxrot2} on the two parameters $\varepsilon$ and $p_0$, is only introduced to first order through the arguments of the exponential functions $\theta_q\left(\varepsilon,p_0\right)$ and $\chi_q\left(\varepsilon,p_0\right)$. In contrast, to first order, the $A_q$ and $B_q$ terms are independent of $\varepsilon$ and $p_0$.
In the absence of initial momentum ($p_0=0$), we may explicitly evaluate $\left.\partial\theta_q/\partial \varepsilon\right|_{\varepsilon=0}$ and $\left.\partial\chi_q/\partial \varepsilon\right|_{\varepsilon=0}$ as outlined in \cite{mcdowall2009fidelity}:
\begin{eqnarray}\label{eq:thetaLumpA}
\left.\frac{\partial\theta_{q}}{\partial\varepsilon}\right|_{\varepsilon=0}& = & \kappa^{2}\frac{\hbar}{2m}\left[\frac{1}{6}\left(N-\frac{1}{N}\right)q-\frac{1}{6}\phi_{d}\left(N^{2}-1\right)\frac{\mathrm{{J}}_{q-1}(N\phi_{d})}{\mathrm{{J}}_{q}(N\phi_{d})}-\left(\frac{1}{3}N+\frac{1}{2}+\frac{1}{6}\frac{1}{N}\right)q^{2}\right],\\\label{eq:chiLumpA}
\left.\frac{\partial\chi_{q}}{\partial\varepsilon}\right|_{\varepsilon=0}& = & -\kappa^{2}\frac{\hbar}{2m}\left[\frac{1}{6}\left(N-\frac{1}{N}\right)q-\frac{1}{6}\phi_{d}\left(N^{2}-1\right)\frac{\mathrm{{J}}_{q-1}(N\phi_{d})}{\mathrm{{J}}_{q}(N\phi_{d})}-\left(\frac{1}{3}N-\frac{1}{2}+\frac{1}{6}\frac{1}{N}\right)q^{2}\right].
\end{eqnarray}
Equations \eqref{eq:thetaLumpA} and \eqref{eq:chiLumpA} differ in the last terms (with $q^2$ dependence). This difference arises due to there being only one free space evolution of duration $T$ between the two pulse sequences leading to $\hat{U}$ ending with a free-space evolution, whereas $\hat{V}^\dagger$ ends with a pulse. Had the free space evolution between the two pulse sequences been of duration $2T$ then Eq. \eqref{eq:thetaLumpA} and \eqref{eq:chiLumpA} would have been equal.
\par Using Eq. \eqref{eq:Ioutcoeff}, and Eqs. (\ref{eq:cncplxrot2}-\ref{eq:chiLumpA}) provides a convenient and accurate way to compute the interferometer output for small $\varepsilon$.

\subsection{Asymptotic Behavior for $\varepsilon \neq 0$}
To understand the general trends of how the interferometer output behaves as function of relevant parameters we now present simple expressions for their asymptotic behavior. We obtain these by keeping only the dominant term in Eq. \eqref{eq:thetaLumpA} and Eq. \eqref{eq:chiLumpA} in the large $N$ limit. For large $N$, $q$ can also be assumed large so we keep only the term proportional to $q^2 N$ which permits us to write: 
\begin{eqnarray}\label{eq:cnCoefepsasym}
c_{q}(\varepsilon) & \simeq & \mathrm{{J}}_{q}(N\phi_{d})\exp\left(i\left(\frac{-\kappa^{2}\hbar Nq^{2}}{6m}\varepsilon-\frac{q\pi}{2}\right)\right),\\\label{eq:dnstrCoefepsasym}
d_{q}^*(\varepsilon) & \simeq & \mathrm{{J}}_{q}(N\phi_{d})\exp\left(i\left(\frac{-\kappa^{2}\hbar Nq^{2}}{6m}\varepsilon+\frac{q\pi}{2}\right)\right).
\end{eqnarray}
Inserting Eq. \eqref{eq:cnCoefepsasym} and Eq. \eqref{eq:dnstrCoefepsasym} into Eq. \eqref{eq:Ioutcoeff}, replacing the quantum momentum distribution with its classical analogue, and taking the continuum limit of the sum, leads to \cite{mcdowall2009fidelity}:
\begin{equation}\label{eq:intoutepsappr}
I(\varepsilon,0)\simeq\mathrm{J}_{0}^{2}\left(N^{3}\phi_{d}{}^{2}\frac{\hbar \kappa^{2}\varepsilon}{6m}\right).
\end{equation}
The oscillatory behavior of the quantum momentum distribution observed in Fig. \ref{fig:pPopViz}(a) for a given $n$ is not captured by its classical counterpart. However, the replacement is valid inasmuch as significant deviation of $I(\varepsilon,0)$ from $0$ requires $\theta_q$ and $\chi_q$ to vary slowly with $q$ and hence the oscillatory behavior of the quantum momentum distribution may be neglected. The full-width at half maximum (FWHM) of the peak of $I(\varepsilon,0)$ about $\varepsilon = 0$ provides a measure of the sensitivity of the interferometer for measuring $T_T$. Equation \eqref{eq:intoutepsappr} show that this width scale as $1/(N^3\phi_d^2)$ in the asymptotic limit. As $N$ defines the interrogation time this scaling shows ``sub-Fourier'' sensitivity of the interferometer. Recall that physically the Fourier inequality limits the minimum width of a measured resonance line $\Delta f$, to the inverse of the time duration $\Delta t$ of the experiment \cite{hecht2001optics}. By instead comparing a higher harmonic of the resonance line then the limiting factor will change in inverse proportion to the number of the harmonic being compared. 
This has been experimentally demonstrated for various systems, an example of which is multi-photon transitions \cite{subfourier18}, where the FWHM of the $q$-th multi-photon line is $\Delta f \Delta t=1/q$. The sub-Fourier narrowing in that system is due to comparison of the $q$-th harmonic of the external driving frequency to the atomic frequency and not the driving frequency itself. Early related work used a quasiperiodically $\delta$-kicked rotor to distinguish between frequencies with sub-Fourier precision \cite{szriftgiser2002observation}. In the present work as well as in \cite{sfcharexperTaluk2010, mcdowall2009fidelity, Horne2011}, the sub-Fourier sensitivity occurs due to the comparison of the frequency at which the pulses are applied (around $1/T_T$) to the frequency at which momentum eigenstates accumulate relative phase between pulses. As the atomic state is a superposition of momentum eigenstates of the form $\left|q \hbar \kappa\right>$, phase accumulation occurs for all involved momentum eigenstates at high harmonics of the recoil frequency. The typical momentum reached in the interferometer increases linearly with $N$ (see Fig. \ref{fig:pPopViz}), and the rate of phase accumulation is proportional to the square of the momentum. For an overall time evolution of duration proportional to $N$, this implies the width of the peak around $\varepsilon = 0$ is proportional to $1/N^3$, as we indeed observed from Eq. \eqref{eq:intoutepsappr}. 

\subsection{Nonzero initial momentum, $p_0 \neq 0$}
We may also analyze the case where the kicking is at resonance ($\varepsilon=0$) and examine the effect of deviation from an initial zero momentum eigenstate. In that case, the terms $\left.\partial \theta_q/\partial p_0\right|_{p_0=0}$ and $\left.\partial \chi_q/\partial p_0\right|_{p_0=0}$ in Eq. (\ref{eq:cncplxrot2}) and (\ref{eq:dncplxrot2}) govern the functional dependence of $c_q\left(0,p_0\right)$ and $d_q^*\left(0,p_0\right)$ on $p_0$ respectively. Following \cite{mcdowall2009fidelity}, we find:
\begin{eqnarray}\label{eq:thetaLumpB}
\left.\frac{\partial\theta_{q}}{\partial p_0}\right|_{p_0=0}&=&-T_T \frac{\kappa}{2m} q(N+1),\\\label{eq:chiLumpB}
\left.\frac{\partial\chi_{q}}{\partial p_0}\right|_{p_0=0}&=&-T_T \frac{\kappa}{2m} q(N-1).
\end{eqnarray}
Using these expressions and Graf's identity for the subsequent evaluation of Eq. \eqref{eq:Ioutcoeff} allows us to write:
\begin{equation}\label{eq:intoutp0appr}
I(0,p_{0})\simeq\mathrm{J}_{0}^{2}\left(N\phi_{d}\sqrt{2-2\cos\left(N\kappa T_{T} p_0/m\right)}\right).
\end{equation}
To obtain a simple relation for the scaling of the width we perform a Taylor series expansion in the argument of the Bessel function to lowest order in $p_0$:
\begin{equation}\label{eq:intoutp0appr2}
I(0,p_0)\simeq\mathrm{J}_{0}^{2}\left(N^2 \phi_d \frac{\kappa  T_T p_0}{m}  \right)
\end{equation}
This shows that to leading order Eq. \eqref{eq:intoutp0appr} exhibits a $1/(N^2\phi_d)$ scaling in the width of the output peak with respect to deviation in initial momentum. This is similar to \cite{sfcharexperTaluk2010, mcdowall2009fidelity} (differing only by a factor of two) and more sensitive than the $1/N$ scaling associated with the quantum resonances of \cite{qres2003,Ryu2006}. The kick sequence can therefore be utilized as a narrow velocity filter \cite{Martin2008, Ullah2011}.

\section{Acceleration}\label{qdka}
In this section we consider the potential sensitivity of the proposed interferometer when used to measure an acceleration. A concrete example would be a measurement of the local gravitational acceleration, by aligning the two counter-propagating laser beams vertically. One can then accelerate the standing wave by chirping one of the laser beam frequencies in a manner that seeks to match the free-falling frame.
 \par To analyze this scenario, we consider the introduction of a constant acceleration $a$ to the Hamiltonian of Eq. \eqref{eq:hamdeltakick}:
\begin{equation}\label{eq:qdkaHam}
\hat{H}_{\delta ka}=\frac{\hat{p}^2}{2m}+ma\hat{x}+\hbar\phi_{d}\cos\left(\kappa\hat{x}\right)\left(\sum_{n=0}^{N-1}\delta\left(t-nT\right)-\sum_{n=N}^{2N-1}\delta(t-nT) \right).
\end{equation}
During free-fall, time-evolution is governed by:
\begin{equation}
\hat{F}(t)=\exp\left(-\frac{i}{\hbar}\left[\frac{\hat{p}^2}{2m}+ma\hat{x} \right]t\right),
\end{equation}
which allows us to express the stroboscopic time evolution operator $\hat{U}$ describing a kick and subsequent evolution in the linear potential for one period $T$ as:
\begin{equation}\label{opsolqdka}
\hat{U}(T) = \hat{F}(T) \exp\left(-i\phi_{d}\cos\left(\kappa\hat{x}\right)\right).
\end{equation}
The operator that describes time evolution when a $\pi$-phase shift of the potential has been made is:
\begin{equation}\label{opsolqdkapi}
\hat{V}(T)  = \hat{F}(T)\exp\left(i\phi_{d}\cos\left(\kappa\hat{x}\right)\right).
\end{equation}
\par We now restrict our attention to $\varepsilon=0$, $p_0=0$. The ASKRS of the proposed interferometer will evolve an initial zero-momentum eigenstate to a final time $2NT_T$. When used for measurements of an acceleration $a$ with respect to an applied optical potential the output will be given by the expression:
\begin{equation}\label{eq:intoutputAccel}
I(a)=\left|\left<0\right|\hat{F}^\dagger(2NT_T)\hat{V}^N\hat{U}^N\left|0\right>\right|^{2}
\end{equation}
Inclusion of the linear term $ma\hat{x}$ in Eqs. (\ref{eq:qdkaHam}--\ref{eq:intoutputAccel}) breaks spatial periodicity and renders Bloch's theorem inapplicable. Quasimomentum conservation may however be restored by performing gauge transformations of Eq. \eqref{eq:qdkaHam} and the states that take part in the dynamics \cite{fishman2003theory, bach2005quantum}. 
In order to simplify the expressions we now proceed by constructing and then evaluating Eq. \eqref{eq:intoutputAccel} in the gauge transformed frame.
This is accomplished under application of the unitary operator $\hat{\mathcal{U}}(t)=\exp\left(-\frac{i}{\hbar}ma\hat{x} t\right)$, which yields $\widetilde{H}=\hat{\mathcal{U}}^\dagger(t)\hat{H}_{\delta k a}\hat{\mathcal{U}}(t)-ma\hat{x}$ (see appendix for details). Explicitly: 
\begin{equation}\label{eq:gtransham}
\widetilde{H}=\frac{1}{2m}\left(\hat{p}-mat\right)^{2}+\hbar\phi_{d}\cos\left(\kappa\hat{x}\right)\left(\sum_{n=0}^{N-1}\delta\left(t-nT_T\right)-\sum_{n=N}^{2N-1}\delta(t-nT_T) \right).
\end{equation}
To facilitate analysis of the states that take part in the dynamics we introduce the time-label $\left|n\hbar\kappa,t=0\right>:=\left|n\hbar\kappa\right>$. A free-fall of duration $2NT_T$ will have a final state given by $\left|n\hbar\kappa,t=2NT_T\right>=\hat{F}(2NT_T)\left|n\hbar\kappa\right>$, where we retain the initial momentum in the label.
\par We now view a state vector $\left|0,t\right>$ as a lab frame state and introduce the transformed frame $\widetilde{\left|0 , t \right>} \equiv \hat{\mathcal{U}}^\dagger(t)\left|0,t\right>$.
The transformation $\hat{H}_{\delta ka}\rightarrow\widetilde{H}$ implies:
\begin{eqnarray}\label{eq:transffket}
\widetilde{\left|0,t\right>} & = & \mathcal{U}^\dagger(t)\left|0,t\right>,\\\label{eq:expabchProd}
& = & \mathcal{U}^\dagger(t)\hat{F}(t) \left|0\right>, \\
& = & \exp\left(-\frac{ima^2t^3}{6\hbar} \right)\exp\left(-\frac{i}{\hbar}\left[\frac{\hat{p}^2}{2m}-\frac{at\hat{p}}{2} \right]t \right)\left|0\right>,\\
& = & \exp\left(-\frac{ima^2t^3}{6\hbar}\right)\left|0\right>,
\end{eqnarray}
where $\exp\left(-ima^2t^3/(6\hbar)\right)$ is a phase term independent of momentum. 
Here and in the sequel, exponential products such as $\mathcal{U}^\dagger(t)\hat{F}(t)$ of Eq. \eqref{eq:expabchProd} will be expanded using the Zassenhaus lemma to the Baker-Campbell-Hausdorff theorem \cite{magnus1954exponential}.
\par
In the transformation $\hat{H}_{\delta k a} \rightarrow \widetilde{H}$ temporal quasi-periodicity is broken, which implies that upon transformation of the kick to kick operator $\hat{U}$ and its $\pi$-shifted counterpart $\hat{V}$, an explicit time-dependence is introduced according to $\widetilde{U}_n=\hat{\mathcal{U}}^\dagger(nT_T)\hat{U}\hat{\mathcal{U}}([n-1]T_T)$. The index $n$ is now defined to be an integer equal to the current kick number, started at one for each kicked rotor sequence. An evaluation of the transformation results in:
\begin{equation}\label{eq:opTrans}
\widetilde{U}_{n}(T_T) = \exp\left(-\frac{i}{\hbar}\left[\frac{\hat{p}^{2}}{2m}T_T-\frac{a}{2}\hat{p}T_T^2(2n-1)\right]\right)\exp\left(-i\phi_{d}\cos\left(\kappa\hat{x}\right)\right).
\end{equation}
Similarly, taking into account the time accrued due to application of the $N$ transformed operators $\widetilde{U}_n$ yields the $\pi$-shifted operators:
\begin{equation}\label{eq:opTransShifted}
\widetilde{V}_{n}(T_T) = \exp\left(-\frac{i}{\hbar}\left[\frac{\hat{p}^{2}}{2m}T_T-\frac{a}{2}\hat{p}T_T^2(2(N+n)-1)\right]\right)\exp\left(i\phi_{d}\cos\left(\kappa\hat{x}\right)\right).
\end{equation}
In both Eq. \eqref{eq:opTrans} and Eq. \eqref{eq:opTransShifted} we omit a global phase term dependent on the index $n$.
Equations (\ref{eq:transffket}--\ref{eq:opTransShifted}) 
imply that Eq. \eqref{eq:intoutputAccel} may be rewritten in the form:
\begin{equation}\label{eq:outputtransformed}
I(a)=\left|\left<0\right| 
\widetilde{V}_N(T_T)\dots \widetilde{V}_1(T_T)
\widetilde{U}_N(T_T)\dots\widetilde{U}_1(T_T)
\left|0\right>\right|^{2},
\end{equation}
where we have used the coincidence of the two frames at $t=0$.
\par Having restored spatial periodicity in Eq. \eqref{eq:outputtransformed}, which in turn renders Bloch's theorem applicable \cite{bach2005quantum, fishman2002stable, ashcroftss1976} we are now in a position to proceed as in section \ref{qdkr}. 
First, write Eq. \eqref{eq:outputtransformed} as a sum over expansion coefficients
\begin{equation}\label{eq:ioAco}
I(a)=\left|\sum_{q\in\mathbb{Z}}d_q^*(a)c_q(a)\right|^2.
\end{equation}
In order to find $c_q(a)$ and $d_q^*(a)$ we require the following expressions for the identity operator: 
\begin{eqnarray}
\hat{\mathbb{I}} & = & \sum_{q\in\mathbb{Z}} {\widetilde{\left|q\hbar\kappa, t \right>}\widetilde{\left< q\hbar\kappa , t \right|}}=\sum_{q\in\mathbb{Z}}\mathcal{U}^\dagger(t)\hat{F}(t)\left|q\hbar\kappa\right>\left<q\hbar\kappa\right|\hat{F}^\dagger(t)\mathcal{U}(t)=\sum_{q\in\mathbb{Z}}\left|q\hbar\kappa\right>\left<q\hbar\kappa\right|, \label{eq:identity}
\end{eqnarray}
which are valid when the initial state has $p_0=0$.
After inserting Eq. \eqref{eq:identity} into Eq. \eqref{eq:outputtransformed} we arrive at:
\begin{align}
c_q(a) &=\left<q\hbar\kappa \right| \widetilde{U}_N(T_T)\dots\widetilde{U}_1(T_T) \left|0 \right>,\\
d_q^*(a) & =\left<0 \right| \widetilde{V}_N(T_T)\dots\widetilde{V}_1(T_T) \left|q\hbar\kappa \right>.
\end{align}
Approximate expressions for small $a$ can be found in a similar manner to previous sections:
\begin{eqnarray}\label{eq:accelCoe}
c_q(a) & \simeq & \mathrm{J}_q(N\phi_d)\exp\left(-i\left(T_T^2 \frac{\kappa}{12} q(N+1)(4N-1)a+q\frac{\pi}{2} \right)\right),\\\label{eq:accelCoeD}
d_q^*(a) & \simeq &\mathrm{J}_q(N\phi_d)\exp\left(-i\left(T_T^2 \frac{\kappa}{12} q(N-1)(8N-1)a-q\frac{\pi}{2} \right)\right).
\end{eqnarray}
Using Eqs. (\ref{eq:ioAco}), (\ref{eq:accelCoe}), and (\ref{eq:accelCoeD}) together with Graf's identity allows one to show that the interferometer output to first order in $a$ is given by:
\begin{equation}\label{eq:intoutputaccel}
I(a)=\mathrm{J}_{0}^{2}\left(N\phi_{d}\sqrt{2-2\cos\left(N\left(2N-1\right) \frac{\kappa T_{T}^{2} a}{2} \right)}\right).
\end{equation}
Expanding the argument of the Bessel function to lowest order in $a$ yields:
\begin{equation}\label{eq:intoutputaccel2}
I(a)=\mathrm{J}_{0}^{2}\left(N^2 (2N-1)\phi_d \frac{a T_T^2\kappa}{2}\right).
\end{equation}
Equation \eqref{eq:intoutputaccel2} exhibits a $1/(N^3 \phi_d)$ scaling for large $N$ in the width of the output peak when $a$ is swept across the resonant zero.
\par
\begin{figure}
\includegraphics[width=8.6cm]{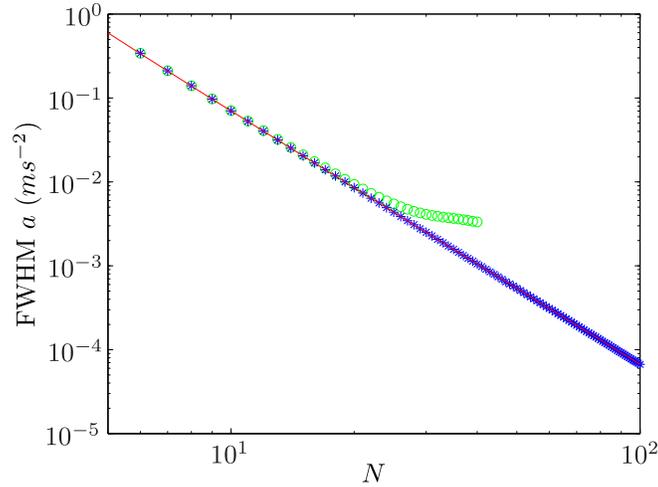} 
\caption{(Color online) FWHM of the interferometer output peak under introduction of a linear acceleration. Red line: Asymptotic behavior for an initial zero-momentum eigenstate $\left|0\right>$ as predicted by Eq. \eqref{eq:intoutputaccel}; Blue stars: Full quantum calculation for an initial state $\left|0\right>$ based on Eq. \eqref{eq:gtransham}; Green circles: Initial Gaussian wavepacket with $\sigma=100\mu m$. Parameters: $\lambda=780nm$, ${}^{85}M_\mathrm{Rb}$, $\phi_d=0.5$.} 
\label{fig:qdkafig}
\end{figure}

In order to check the veracity of the above calculation Fig. \ref{fig:qdkafig} compares the FWHM as predicted by Eq. \eqref{eq:intoutputaccel} and a numerical calculation based on Eq. \eqref{eq:outputtransformed}. Good agreement is seen over the range of parameters investigated. 

The asymptotic behaviours ($1/(N^3 \phi_d)$ for measurements of accelerations and $1/(N^3\phi_d^2)$ for measurements of the Talbot time) are derived under the assumption that the initial state is $\left| 0 \right\rangle $. As any experiment is limited by finite size and temperature, we numerically inspect the behavior of the system for an initial Gaussian wavefunction with width $\sigma=100\mathrm{\mu m}$. This is done for realistic experimental parameters: the mass of the atoms is taken to be that of ${}^{85}\mathrm{Rb}$; the effective potential $\phi_d=0.5$ and the standing wave formed by laser-light of $\lambda=780\mathrm{nm}$. The results also shown in Fig. \ref{fig:qdkafig} agrees well with Eq. \eqref{eq:intoutputaccel} for small $N$, but for $N>25$ significant deviation from the $1/(N^3 \phi_d)$ scaling law is seen. The deviation can be understood in terms of Eq. \eqref{eq:intoutp0appr2}. When the momentum width of the initial state is much smaller than the width of the output peak (of Eq. \eqref{eq:intoutp0appr2}), we expect $\left| 0 \right\rangle $ to approximate the initial state well and the asymptotic results to be valid. However, as $N$ is increased the output peak narrows. When this peak becomes comparable to the momentum width of the initial state, we would indeed expect to see deviations from the scaling laws. As pointed out in \cite{Horne2011} extension of the regime with favorable scaling requires narrow initial momentum distributions. In order for the interferometer to be competitive for precision measurements new atomic sources may therefore be needed.

\section{Finite pulse duration}\label{sec:longpulse}
In previous sections we analyzed the short pulse limit. This is beneficial as it allows for analytical predictions that enable a detailed understanding of the underlying principles of the interferometer. An actual implementation requires finite pulse durations. One reason for this is the finite laser power available for generation of the optical standing wave. Moreover, investigations of other related atom interferometers have shown that violating the Raman-Nath limit can improve the sensitivity \cite{andersen2009lattice}. In this section we therefore investigate the performance of the interferometer for finite pulse durations $\tau_{p}$.
Because significant dynamics occur during interaction with the optical standing wave, we are required to retain the $\hat{p}^2/2m$ term in the Hamiltonian during the pulse. The one-period pulse to pulse evolution operator is now given by:
\begin{equation} \label{oplp}
 \hat{U}=\exp\left(-\frac{i}{\hbar}\frac{\hat{p}^{2}}{2m}\left( T-\tau_p \right)\right)\exp\left( -\frac{i}{\hbar}\left( \frac{\hat{p}^{2}}{2m} + \frac{\mathcal{V}_0}{2} \cos\left(\kappa \hat{x}\right) \right)\tau_p \right)
\end{equation}
The strength of the optical potential is now written as $\mathcal{V}_0$.
Equation \eqref{oplp} can be rewritten using the Zassenhaus lemma \cite{magnus1954exponential}: 
\begin{eqnarray}\label{eq:finopexp}
\hat{U} & = & \exp\left(-\frac{i}{\hbar}\frac{\hat{p}^{2}}{2m}T\right)\exp\left(-\frac{i}{\hbar}\frac{\mathcal{V}_0}{2}\cos\left(\kappa\hat{x}\right)\tau_{p}\right)\nonumber\\
 &  & \times\exp\left(-\frac{\tau_{p}^{2}}{2}\left[-\frac{i}{\hbar}\frac{\hat{p}^{2}}{2m},-\frac{i}{\hbar}\frac{\mathcal{V}_0}{2}\cos\left(\kappa\hat{x}\right)\right]\right)\nonumber\\
 &  & \times\exp\left(\mathcal{O}\left(\tau_p^3\right)\right)\dots\end{eqnarray}
For any state $\left| \beta=0 \right\rangle$ that can be expressed as $\left| \beta=0 \right\rangle =\sum_{q\in \mathbb{Z}} b_q \left| q \hbar \kappa \right\rangle$ we have $\exp\left(-\frac{i}{\hbar}\frac{\hat{p}^{2}}{2m}T_{T}\right) \left| \beta=0 \right\rangle =\left| \beta=0 \right\rangle$. Hence, for $T=T_T$:
\begin{eqnarray}
\hat{U}\left| \beta=0 \right\rangle & = &  \exp\left(-\frac{i}{\hbar}\frac{\mathcal{V}_0}{2}\cos\left(\kappa\hat{x}\right)\tau_{p}\right)\nonumber\\
 &  & \times\exp\left(-\frac{\tau_{p}^{2}}{2}\left[-\frac{i}{\hbar}\frac{\hat{p}^{2}}{2m},-\frac{i}{\hbar}\frac{\mathcal{V}_0}{2}\cos\left(\kappa\hat{x}\right)\right]\right)\nonumber\\ \label{eq:handwaving}
 &  & \times\exp\left(\mathcal{O}\left(\tau_p^3\right)\right)\ldots  \left| \beta=0 \right\rangle \end{eqnarray} 
The first term in this expression is the interaction with the potential that in the short pulse limit gives rise to unbounded ballistic transfer of momentum to the atoms at quantum resonance. For finite $\tau_p$, the subsequent terms containing commutators, which arise due to dynamics of the atoms during the pulse, are expected to obstruct this transfer. Note that the lowest order in which $\tau_p$ appears in these terms is quadratic.

Cancellation of a term linear in $\tau_p$ is the advantage of using two kicked rotor pulse sequences over simply using two single pulses as in \cite{temporalTalbot}. In order to compare to a single pulse of duration $\tau_p$,  we rewrite the time evolution operator:
\begin{eqnarray}\label{eq:singlep}
\hat{U}_{\mathrm{sp}} \left| \beta=0 \right\rangle & = & \exp\left(-\frac{i}{\hbar}\frac{\mathcal{V}_0}{2}\cos\left(\kappa\hat{x}\right)\tau_{p}\right) 
 \nonumber\\
 &  & \times \exp\left(-\frac{i}{\hbar}\frac{\hat{p}^{2}}{2m} \tau_p \right) \nonumber\\
 &  & \times\exp\left(\mathcal{O}\left(\tau_p^2\right)\right)\ldots \left| \beta=0 \right\rangle \end{eqnarray}
which has a term arising from the dynamics of the atoms that is linear in $\tau_p$. For a single pulse the energy that can be transferred to an atom initially at rest is bounded by the depth of the standing wave potential. This limitation in transferred energy and thereby in $\Delta p$ may be overcome using a kicked rotor train of pulses of finite duration as the first order term arising from dynamics of the atoms in the optical standing wave cancels. 

\begin{figure}
\includegraphics[width=8.6cm]{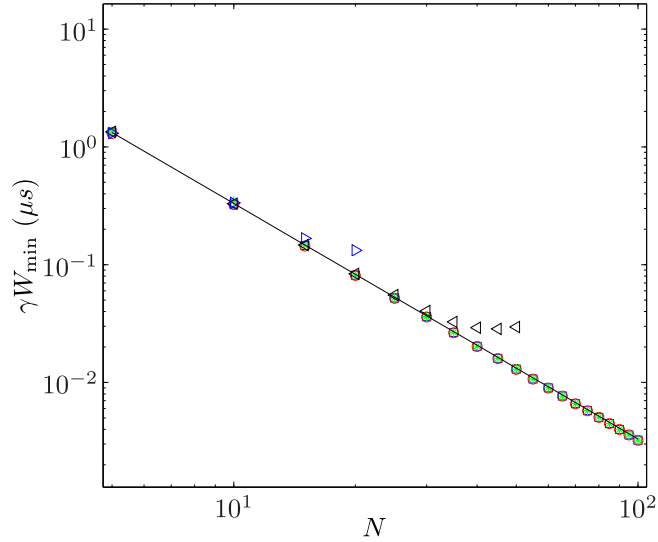} 
\caption{(Color online) Minimum FWHM under a deviation $\varepsilon$: $W_{\mathrm{min}}$ (scaled by $\gamma$), numerically computed as described in the text. Results for an initial state $\left|0\right>$ are shown for potential strengths of: $\gamma=1$ $(\square)$, $\gamma=10$ $(\bigcirc)$, $\gamma=100$ $(\ast)$. Results for an initial Gaussian state $\left|G\right>$ with $\sigma=100\mu m$ are shown for potential strengths: $\gamma=1$ $(\triangleleft)$, $\gamma=10$ $(\triangleright)$. Equation \eqref{wmin} is shown as the black line. Parameters: $\lambda=780nm$, ${}^{85}M_\mathrm{Rb}$.}
\label{fig:finfwhmData}
\end{figure}

For an initial zero momentum eigenstate the interferometer output is still given by Eq. \eqref{eq:interoutput}, with $\hat{U}$ given by Eq. \eqref{oplp} and $\hat{V}$ defined similarly with the phase of the potential shifted by $\pi$. In order to investigate the potential sensitivity of the interferometer for measuring the Talbot time we set $p_0=0$ and numerically compute the interferometer output for sets of parameters $N$, $\mathcal{V}_0$, and $\tau_p$ while scanning $T$ across $T_T$. Similarly to the short pulse case we observe a resonant peak in the interferometer output for $T \approx T_T$. The FWHM of this peak is a measure of the potential sensitivity of the interferometer. For each combination of $N$ and $\mathcal{V}_0$ we compute this width while increasing $\tau_p$ from zero until a minimum in the width is found. Define $\gamma=m \mathcal{V}_0/\left( \hbar\kappa \right)^2$; a measure of the potential strength in units of the energy transfer associated with a two-photon process. Figure \ref{fig:finfwhmData} shows the resulting minimum width $W_\mathrm{min}$ multiplied by $\gamma$ as a function of $N$ plotted on a double log scale. We observe a remarkably simple relation as all points from computations taking the initial state as $\left| 0 \right\rangle$ approximately falls on a straight line with slope -2. A simple relation shown as a straight line in Fig. \ref{fig:finfwhmData} therefore predicts the $W_\mathrm{min}$ over a large range of parameters:
\begin{equation}\label{wmin}
 W_\mathrm{min} \approx \frac{33~\mu s}{\gamma N^2}
\end{equation}
where the $33~\mu s$ is found by fitting.

\begin{figure}
\includegraphics[width=8.6cm]{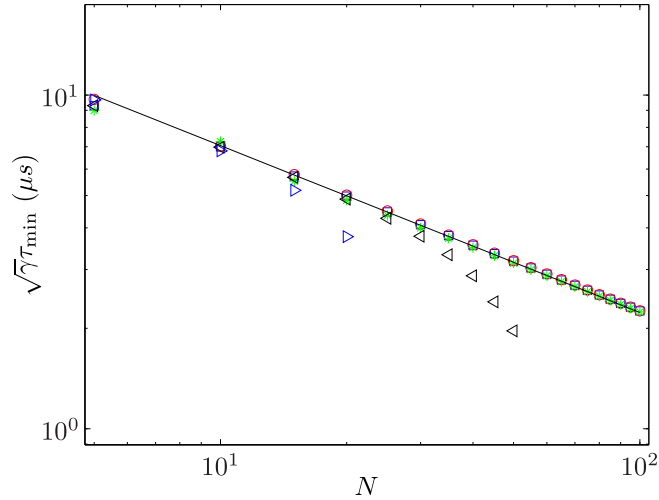} 
\caption{(Color online) Pulse duration (scaled by $\sqrt{\gamma}$) required to minimize FWHM as a function of $N$. Results for an initial state $\left|0\right>$ is shown for potential strengths: $\gamma=1$ $(\square)$, $\gamma=10$ $(\bigcirc)$, $\gamma=100$ $(\ast)$. Results for an initial Gaussian state $\left|G\right>$ with $\sigma=100\mu m$ are shown for potential strengths: $\gamma=1$ $(\triangleleft)$, $\gamma=10$ $(\triangleright)$. Equation \eqref{taumin} is shown as the black line. Parameters: $\lambda=780nm$, ${}^{85}M_\mathrm{Rb}$.}
\label{fig:fintauData}
\end{figure}

We denote the pulse duration that yields $W_{min}$ for given $N$ and $\mathcal{V}_0$ for $\tau_{min}$. In Fig. \ref{fig:fintauData} we plot $\sqrt{\gamma} \tau_{min}$ as a function of $N$ and again observe that all points from an initial state of $\left| 0 \right\rangle$ follow the simple relation that is shown as a line: 
\begin{equation}\label{taumin}
 \tau_{min} \approx \frac{22~\mu s}{\sqrt{\gamma N}}
\end{equation}

Using a typical momentum of an atom after the first kicked rotor pulse sequence we find that atoms move on the order the period of the optical standing wave during a pulse duration $\tau_{min}$, independent of $N$ and $\gamma$. We therefore observe that the ``optimal'' pulse duration $\tau_{min}$ is not within the Raman-Nath regime but occurs upon moderate violation thereof.
These scaling relations are verified for a wide range of $\gamma$ and $N$. When the initial state is taken to be $\left| 0 \right\rangle $, then for values of $\gamma N$ much larger than unity, the behavior of the system is well captured by the simple relations of Eqs. \eqref{wmin} and \eqref{taumin} (see Fig. \ref{fig:finfwhmData} and \ref{fig:fintauData}). For $\gamma N<1$ the system does not display the behavior outlined above. This is due to the reduction to an effective three-level system consisting of $\left| 0 \right\rangle $ and $\left| \pm \hbar \kappa \right\rangle $ where qualitatively different dynamics occur.

Equation \eqref{wmin} show that the width of the peak close to $T=T_T$ only decrease as $1/N^2$ whereas it was $1/N^3$ in the Raman-Nath regime. This is due to the fact that the optimal pulse duration is decreased with $N$ (see Eq. \eqref{taumin}) thereby effectively reducing the kick strength as $N$ is increased. It should be noted that given $\mathcal{V}_0$ and $N$ choosing $\tau$ according to Eq. \eqref{taumin} will give a much narrower peak and a more sensitive interferometer than choosing a smaller $\tau$ that would fulfill the Raman-Nath condition for all pulses.

Upon extending our numerical calculation to an initial Gaussian wave-packet $\left|G\right>$ (using the Fourier split-step operator method \cite{Wang200517}) we find good agreement for low $N$ (see Fig. \ref{fig:finfwhmData} and \ref{fig:fintauData}), however significant deviation occurs as $N$ is further increased. We ascribe this to the non-zero momentum spread in the Gaussian wave-packet. 

We also note one other feature observed, the output profile generated when modelling the effect of an introduction of $\varepsilon$ was not centered exactly on the $T=T_T$ but displayed a slight shift $\delta \varepsilon$. For accurate measurements this shift could introduce unwanted errors when measuring $T_T$. However, up to the numerical accuracy of our calculations $\delta \varepsilon$ remained unchanged when considering the peaks centered close to $T$ being integer multiples of $T_T$. Systematic effects due to the shift $\delta \varepsilon$ can therefore be reduced by measuring the difference between the position of the peak close to $T=T_T$ and one at a higher multiple of the Talbot time \cite{andersen2009lattice}.

\section{Conclusion}
We have investigated a scheme for an atom interferometer, that generates high momentum differences between its arms through consecutive low order diffraction processes. The interferometer can potentially measure the Talbot time, an initial momentum of the atoms, or accelerations. It builds on quantum resonances in the atom optics $\delta$-kicked rotor and we presented approximate analytical expressions for the sensitivity of operation in the vicinity of quantum resonance. These gave simple scaling relations for the width of the interferometer output peak in the large $N$ limit, with $2N$ being the total interrogation time.

We numerically explored the finite pulse-duration regime, finding relations that predict the optimal pulse duration (not in the Raman-Nath regime) and the sensitivity of the interferometer. The interferometer may be of interest for measurements of the local gravitational acceleration or the fine structure constant $\alpha$.

\section{Appendix} 
\label{gaugexFormapp}
A Hamiltonian that describes a particular system may be transformed to a convenient frame in order to simplify calculations \cite{bach2005quantum, sakurai1994modern}. In this section we outline the gauge-transformation procedure, making explicit the connection between the gauge-transformed Hamiltonian denoted as $\widetilde{H}$ and its ``lab-frame'' counterpart $\hat{H}$. Given a system described by the Hamiltonian $\hat{H}$, then the time-evolution of an initial state $\left|\psi\right>$, can be represented in the Schr\"odinger picture as
\begin{equation}\label{eq:minischro}
i\hbar\frac{\partial \left|\psi\right>}{\partial t}=\hat{H}\left|\psi\right>.
\end{equation}
We now consider the effect of using the operator $\hat{\mathcal{U}}=\exp\left(-i m a \hat{x}t/\hbar\right)$ to construct the gauge transformed wavefunction $\widetilde{\left|\psi\right>}=\hat{\mathcal{U}}^\dagger\left|\psi\right>$.
Taking the time derivative of $\widetilde{\left|\psi\right>}$ and using Eq. \eqref{eq:minischro} yields:
\begin{eqnarray}
i\hbar\frac{\partial}{\partial t}\left[\hat{\mathcal{U}}^\dagger\left|\psi\right>\right] & = & i\hbar\hat{\mathcal{U}}^\dagger\frac{\partial}{ \partial t}\left|\psi\right>+i\hbar\frac{ \partial \hat{\mathcal{U}}^\dagger}{ \partial t}\left|\psi\right>,\nonumber\\
 & = & \left(\hat{\mathcal{U}}^\dagger\hat{H}\hat{\mathcal{U}}-ma\hat{x}\right)\hat{\mathcal{U}}^\dagger\left|\psi\right>.\end{eqnarray}
This implies that time-evolution of the gauge-transformed state $\widetilde{\left|\psi\right>}$, is governed by the Hamiltonian $\widetilde{H}=\hat{\mathcal{U}}^\dagger\hat{H}\hat{\mathcal{U}}-ma\hat{x}$.

\section{Acknowledgments}
This work is supported by NZ-FRST Contract No. NERF-UOOX0703.


\begin{thebibliography}{35}
\expandafter\ifx\csname natexlab\endcsname\relax\def\natexlab#1{#1}\fi
\expandafter\ifx\csname bibnamefont\endcsname\relax
  \def\bibnamefont#1{#1}\fi
\expandafter\ifx\csname bibfnamefont\endcsname\relax
  \def\bibfnamefont#1{#1}\fi
\expandafter\ifx\csname citenamefont\endcsname\relax
  \def\citenamefont#1{#1}\fi
\expandafter\ifx\csname url\endcsname\relax
  \def\url#1{\texttt{#1}}\fi
\expandafter\ifx\csname urlprefix\endcsname\relax\def\urlprefix{URL }\fi
\providecommand{\bibinfo}[2]{#2}
\providecommand{\eprint}[2][]{\url{#2}}

\bibitem[{\citenamefont{Berman}(1996)}]{berman1996atom}
\bibinfo{author}{\bibfnamefont{P.~R.} \bibnamefont{Berman}},
  \emph{\bibinfo{title}{{Atom Interferometry}}} (\bibinfo{publisher}{Academic
  Press}, \bibinfo{year}{1996}).

\bibitem[{\citenamefont{Cronin et~al.}(2009)\citenamefont{Cronin, Schmiedmayer,
  and Pritchard}}]{cronin2009optics}
\bibinfo{author}{\bibfnamefont{A.}~\bibnamefont{Cronin}},
  \bibinfo{author}{\bibfnamefont{J.}~\bibnamefont{Schmiedmayer}},
  \bibnamefont{and}
  \bibinfo{author}{\bibfnamefont{D.}~\bibnamefont{Pritchard}},
  \bibinfo{journal}{Rev. Mod. Phys.} \textbf{\bibinfo{volume}{81}},
  \bibinfo{pages}{1051} (\bibinfo{year}{2009}).

\bibitem[{\citenamefont{Cadoret et~al.}(2008)\citenamefont{Cadoret,
  de~Mirandes, Clad\'e, Guellati-Kh\'elifa, Schwob, Nez, Julien, and
  Biraben}}]{cadoret2008combination}
\bibinfo{author}{\bibfnamefont{R.}~\bibnamefont{Bouchendira}},
  \bibinfo{author}{\bibfnamefont{P.}~\bibnamefont{Clad\'e}},
  \bibinfo{author}{\bibfnamefont{S.}~\bibnamefont{Guellati-Kh\'elifa}},
  \bibinfo{author}{\bibfnamefont{F.}~\bibnamefont{Nez}}, \bibnamefont{and}
  \bibinfo{author}{\bibfnamefont{F.}~\bibnamefont{Biraben}},
  \bibinfo{journal}{Phys. Rev. Lett.} \textbf{\bibinfo{volume}{106}},
  \bibinfo{pages}{080801} (\bibinfo{year}{2011}).


\bibitem[{\citenamefont{Fixler et~al.}(2007)\citenamefont{Fixler, Foster,
  McGuirk, and Kasevich}}]{fixler2007atom}
\bibinfo{author}{\bibfnamefont{J.}~\bibnamefont{Fixler}},
  \bibinfo{author}{\bibfnamefont{G.}~\bibnamefont{Foster}},
  \bibinfo{author}{\bibfnamefont{J.}~\bibnamefont{McGuirk}}, \bibnamefont{and}
  \bibinfo{author}{\bibfnamefont{M.}~\bibnamefont{Kasevich}},
  \bibinfo{journal}{Science} \textbf{\bibinfo{volume}{315}},
  \bibinfo{pages}{74} (\bibinfo{year}{2007}).

\bibitem[{\citenamefont{Poli et~al.}(2011)\citenamefont{Poli, Wang, Tarallo,
  Alberti, Prevedelli, and Tino}}]{Poli2011}
\bibinfo{author}{\bibfnamefont{N.}~\bibnamefont{Poli}},
  \bibinfo{author}{\bibfnamefont{F.~Y.}~\bibnamefont{Wang}},
  \bibinfo{author}{\bibfnamefont{M.~G.}~\bibnamefont{Tarallo}},
  \bibinfo{author}{\bibfnamefont{A.}~\bibnamefont{Alberti}},
  \bibinfo{author}{\bibfnamefont{M.}~\bibnamefont{Prevedelli}},
  \bibnamefont{and} \bibinfo{author}{\bibfnamefont{G.~M.}~\bibnamefont{Tino}},
  \bibinfo{journal}{Phys. Rev. Lett.} \textbf{\bibinfo{volume}{106}},
  \bibinfo{pages}{038501} (\bibinfo{year}{2011}).

\bibitem[{\citenamefont{Merlet et~al.}(2010)\citenamefont{Merlet, Bodart,
  Malossi, Landragin, Santos, Gitlein, and Timmen}}]{Merlet2010}
\bibinfo{author}{\bibfnamefont{S.}~\bibnamefont{Merlet}},
  \bibinfo{author}{\bibfnamefont{Q.}~\bibnamefont{Bodart}},
  \bibinfo{author}{\bibfnamefont{N.}~\bibnamefont{Malossi}},
  \bibinfo{author}{\bibfnamefont{A.}~\bibnamefont{Landragin}},
  \bibinfo{author}{\bibfnamefont{F.}~\bibnamefont{Santos}},
  \bibinfo{author}{\bibfnamefont{O.}~\bibnamefont{Gitlein}}, \bibnamefont{and}
  \bibinfo{author}{\bibfnamefont{L.}~\bibnamefont{Timmen}},
  \bibinfo{journal}{Metrologia} \textbf{\bibinfo{volume}{47}},
  \bibinfo{pages}{L9} (\bibinfo{year}{2010}).

\bibitem[{\citenamefont{Gupta et~al.}(2002)\citenamefont{Gupta, Dieckmann,
  Hadzibabic, and Pritchard}}]{Gupta2002}
\bibinfo{author}{\bibfnamefont{S.}~\bibnamefont{Gupta}},
  \bibinfo{author}{\bibfnamefont{K.}~\bibnamefont{Dieckmann}},
  \bibinfo{author}{\bibfnamefont{Z.}~\bibnamefont{Hadzibabic}},
  \bibnamefont{and}
  \bibinfo{author}{\bibfnamefont{D.~E.}~\bibnamefont{Pritchard}},
  \bibinfo{journal}{Phys. Rev. Lett.} \textbf{\bibinfo{volume}{89}},
  \bibinfo{pages}{140401} (\bibinfo{year}{2002}).

\bibitem[{\citenamefont{M\"uller
  et~al.}(2008{\natexlab{a}})\citenamefont{M\"uller, Chiow, Long, Herrmann, and
  Chu}}]{muller2008atom}
\bibinfo{author}{\bibfnamefont{H.}~\bibnamefont{M\"uller}},
  \bibinfo{author}{\bibfnamefont{S.-w.} \bibnamefont{Chiow}},
  \bibinfo{author}{\bibfnamefont{Q.}~\bibnamefont{Long}},
  \bibinfo{author}{\bibfnamefont{S.}~\bibnamefont{Herrmann}}, \bibnamefont{and}
  \bibinfo{author}{\bibfnamefont{S.}~\bibnamefont{Chu}},
  \bibinfo{journal}{Phys. Rev. Lett.} \textbf{\bibinfo{volume}{100}},
  \bibinfo{pages}{180405} (\bibinfo{year}{2008}{\natexlab{a}}).

\bibitem[{\citenamefont{Talukdar et~al.}(2010)\citenamefont{Talukdar, Shrestha,
  and Summy}}]{sfcharexperTaluk2010}
\bibinfo{author}{\bibfnamefont{I.}~\bibnamefont{Talukdar}},
  \bibinfo{author}{\bibfnamefont{R.}~\bibnamefont{Shrestha}}, \bibnamefont{and}
  \bibinfo{author}{\bibfnamefont{G.~S.} \bibnamefont{Summy}},
  \bibinfo{journal}{Phys. Rev. Lett.} \textbf{\bibinfo{volume}{105}},
  \bibinfo{pages}{054103} (\bibinfo{year}{2010}).

\bibitem[{\citenamefont{M\"uller
  et~al.}(2008{\natexlab{b}})\citenamefont{M\"uller, Chiow, and
  Chu}}]{mullertheory2008}
\bibinfo{author}{\bibfnamefont{H.}~\bibnamefont{M\"uller}},
  \bibinfo{author}{\bibfnamefont{S.-w.} \bibnamefont{Chiow}}, \bibnamefont{and}
  \bibinfo{author}{\bibfnamefont{S.}~\bibnamefont{Chu}},
  \bibinfo{journal}{Phys. Rev. A} \textbf{\bibinfo{volume}{77}},
  \bibinfo{pages}{023609} (\bibinfo{year}{2008}{\natexlab{b}}).

\bibitem[{\citenamefont{Clad{\'e} et~al.}(2009)\citenamefont{Clad{\'e},
  Guellati-Kh{\'e}lifa, Nez, and Biraben}}]{clade2009large}
\bibinfo{author}{\bibfnamefont{P.}~\bibnamefont{Clad{\'e}}},
  \bibinfo{author}{\bibfnamefont{S.}~\bibnamefont{Guellati-Kh{\'e}lifa}},
  \bibinfo{author}{\bibfnamefont{F.}~\bibnamefont{Nez}}, \bibnamefont{and}
  \bibinfo{author}{\bibfnamefont{F.}~\bibnamefont{Biraben}},
  \bibinfo{journal}{Phys. Rev. Lett.} \textbf{\bibinfo{volume}{102}},
  \bibinfo{pages}{240402} (\bibinfo{year}{2009}).

\bibitem[{\citenamefont{M\"uller et~al.}(2009)\citenamefont{M\"uller, Chiow,
  Herrmann, and Chu}}]{mullerembedded2009}
\bibinfo{author}{\bibfnamefont{H.}~\bibnamefont{M\"uller}},
  \bibinfo{author}{\bibfnamefont{S.-w.} \bibnamefont{Chiow}},
  \bibinfo{author}{\bibfnamefont{S.}~\bibnamefont{Herrmann}}, \bibnamefont{and}
  \bibinfo{author}{\bibfnamefont{S.}~\bibnamefont{Chu}},
  \bibinfo{journal}{Phys. Rev. Lett.} \textbf{\bibinfo{volume}{102}},
  \bibinfo{pages}{240403} (\bibinfo{year}{2009}).

\bibitem[{\citenamefont{Chiow et~al.}(2011)\citenamefont{Chiow, Kovachy, Chien, and Kasevich}}]{Chiow2011}
\bibinfo{author}{\bibfnamefont{S.-w.}~\bibnamefont{Chiow}},
  \bibinfo{author}{\bibfnamefont{T.} \bibnamefont{Kovachy}},
  \bibinfo{author}{\bibfnamefont{H.-C}~\bibnamefont{Chien}}, \bibnamefont{and}
  \bibinfo{author}{\bibfnamefont{M.~A.}~\bibnamefont{Kasevich}},
  \bibinfo{journal}{Phys. Rev. Lett.} \textbf{\bibinfo{volume}{107}},
  \bibinfo{pages}{130403} (\bibinfo{year}{2011}).

\bibitem[{\citenamefont{McDowall et~al.}(2009)\citenamefont{McDowall, Hilliard,
  McGovern, Gr{\"u}nzweig, and Andersen}}]{mcdowall2009fidelity}
\bibinfo{author}{\bibfnamefont{P.}~\bibnamefont{McDowall}},
  \bibinfo{author}{\bibfnamefont{A.}~\bibnamefont{Hilliard}},
  \bibinfo{author}{\bibfnamefont{M.}~\bibnamefont{McGovern}},
  \bibinfo{author}{\bibfnamefont{T.}~\bibnamefont{Gr{\"u}nzweig}},
  \bibnamefont{and} \bibinfo{author}{\bibfnamefont{M.~F.}~\bibnamefont{Andersen}},
  \bibinfo{journal}{New J. Phys.} \textbf{\bibinfo{volume}{11}},
  \bibinfo{pages}{123021} (\bibinfo{year}{2009}).

\bibitem[{\citenamefont{Moore et~al.}(1994)\citenamefont{Moore, Robinson,
  Bharucha, Williams, and Raizen}}]{Moore1994}
\bibinfo{author}{\bibfnamefont{F.~L.} \bibnamefont{Moore}},
  \bibinfo{author}{\bibfnamefont{J.~C.} \bibnamefont{Robinson}},
  \bibinfo{author}{\bibfnamefont{C.}~\bibnamefont{Bharucha}},
  \bibinfo{author}{\bibfnamefont{P.~E.} \bibnamefont{Williams}},
  \bibnamefont{and} \bibinfo{author}{\bibfnamefont{M.~G.}
  \bibnamefont{Raizen}}, \bibinfo{journal}{Phys. Rev. Lett.}
  \textbf{\bibinfo{volume}{73}}, \bibinfo{pages}{2974} (\bibinfo{year}{1994}).

\bibitem[{\citenamefont{Moore et~al.}(1995)\citenamefont{Moore, Robinson,
  Bharucha, Sundaram, and Raizen}}]{Moore1995}
\bibinfo{author}{\bibfnamefont{F.~L.} \bibnamefont{Moore}},
  \bibinfo{author}{\bibfnamefont{J.~C.} \bibnamefont{Robinson}},
  \bibinfo{author}{\bibfnamefont{C.~F.} \bibnamefont{Bharucha}},
  \bibinfo{author}{\bibfnamefont{B.}~\bibnamefont{Sundaram}}, \bibnamefont{and}
  \bibinfo{author}{\bibfnamefont{M.~G.} \bibnamefont{Raizen}},
  \bibinfo{journal}{Phys. Rev. Lett.} \textbf{\bibinfo{volume}{75}},
  \bibinfo{pages}{4598} (\bibinfo{year}{1995}).

\bibitem[{\citenamefont{Raizen}(1999)}]{Raizen1999}
\bibinfo{author}{\bibfnamefont{M.}~\bibnamefont{Raizen}},
  \bibinfo{journal}{Adv. At. Mol. Opt. Phy.} \textbf{\bibinfo{volume}{41}},
  \bibinfo{pages}{43} (\bibinfo{year}{1999}).

\bibitem[{\citenamefont{Deng et~al.}(1999)\citenamefont{Deng, Hagley,
  Denschlag, Simsarian, Edwards, Clark, Helmerson, Rolston, and
  Phillips}}]{temporalTalbot}
\bibinfo{author}{\bibfnamefont{L.}~\bibnamefont{Deng}},
  \bibinfo{author}{\bibfnamefont{E.~W.} \bibnamefont{Hagley}},
  \bibinfo{author}{\bibfnamefont{J.}~\bibnamefont{Denschlag}},
  \bibinfo{author}{\bibfnamefont{J.~E.} \bibnamefont{Simsarian}},
  \bibinfo{author}{\bibfnamefont{M.}~\bibnamefont{Edwards}},
  \bibinfo{author}{\bibfnamefont{C.~W.} \bibnamefont{Clark}},
  \bibinfo{author}{\bibfnamefont{K.}~\bibnamefont{Helmerson}},
  \bibinfo{author}{\bibfnamefont{S.~L.} \bibnamefont{Rolston}},
  \bibnamefont{and} \bibinfo{author}{\bibfnamefont{W.~D.}
  \bibnamefont{Phillips}}, \bibinfo{journal}{Phys. Rev. Lett.}
  \textbf{\bibinfo{volume}{83}}, \bibinfo{pages}{5407} (\bibinfo{year}{1999}).

\bibitem[{\citenamefont{Wimberger et~al.}(2003)\citenamefont{Wimberger,
  Guarneri, and Fishman}}]{qres2003}
\bibinfo{author}{\bibfnamefont{S.}~\bibnamefont{Wimberger}},
  \bibinfo{author}{\bibfnamefont{I.}~\bibnamefont{Guarneri}}, \bibnamefont{and}
  \bibinfo{author}{\bibfnamefont{S.}~\bibnamefont{Fishman}},
  \bibinfo{journal}{Nonlinearity} \textbf{\bibinfo{volume}{16}},
  \bibinfo{pages}{1381} (\bibinfo{year}{2003}).

\bibitem[{\citenamefont{Izrailev and
  Shepelyanskii}(1980)}]{izrailev1980quantum}
\bibinfo{author}{\bibfnamefont{F.}~\bibnamefont{Izrailev}} \bibnamefont{and}
  \bibinfo{author}{\bibfnamefont{D.}~\bibnamefont{Shepelyanskii}},
  \bibinfo{journal}{Theor. Math. Phys.} \textbf{\bibinfo{volume}{43}},
  \bibinfo{pages}{553} (\bibinfo{year}{1980}).

\bibitem[{\citenamefont{Ryu et~al.}(2006)\citenamefont{Ryu, Andersen, Vaziri,
  d'Arcy, Grossman, Helmerson, and Phillips}}]{Ryu2006}
\bibinfo{author}{\bibfnamefont{C.}~\bibnamefont{Ryu}},
  \bibinfo{author}{\bibfnamefont{M.~F.}~\bibnamefont{Andersen}},
  \bibinfo{author}{\bibfnamefont{A.}~\bibnamefont{Vaziri}},
  \bibinfo{author}{\bibfnamefont{M.~B.}~\bibnamefont{d'Arcy}},
  \bibinfo{author}{\bibfnamefont{J.~M.}~\bibnamefont{Grossman}},
  \bibinfo{author}{\bibfnamefont{K.}~\bibnamefont{Helmerson}},
  \bibnamefont{and} \bibinfo{author}{\bibfnamefont{W.~D}~\bibnamefont{Phillips}},
  \bibinfo{journal}{Phys. Rev. Lett.} \textbf{\bibinfo{volume}{96}},
  \bibinfo{pages}{160403} (\bibinfo{year}{2006}).

\bibitem[{\citenamefont{Szriftgiser et~al.}(2002)\citenamefont{Szriftgiser,
  Ringot, Delande, and Garreau}}]{szriftgiser2002observation}
\bibinfo{author}{\bibfnamefont{P.}~\bibnamefont{Szriftgiser}},
  \bibinfo{author}{\bibfnamefont{J.}~\bibnamefont{Ringot}},
  \bibinfo{author}{\bibfnamefont{D.}~\bibnamefont{Delande}}, \bibnamefont{and}
  \bibinfo{author}{\bibfnamefont{J.~C.}~\bibnamefont{Garreau}},
  \bibinfo{journal}{Phys. Rev. Lett.} \textbf{\bibinfo{volume}{89}},
  \bibinfo{pages}{224101} (\bibinfo{year}{2002}).

\bibitem[{\citenamefont{Tonyushkin and Prentiss}(2008)}]{Tonyushkin2008}
\bibinfo{author}{\bibfnamefont{A.}~\bibnamefont{Tonyushkin}}
  \bibnamefont{and} \bibinfo{author}{\bibfnamefont{M.}~\bibnamefont{Prentiss}},
  \bibinfo{journal}{Phys. Rev. A} \textbf{\bibinfo{volume}{78}},
  \bibinfo{pages}{053625} (\bibinfo{year}{2008}).

\bibitem[{\citenamefont{Horne et~al.}(2011)}]{Horne2011}
\bibinfo{author}{\bibfnamefont{R.~A.}~\bibnamefont{Horne}}, 
\bibinfo{author}{\bibfnamefont{R.~H..}~\bibnamefont{Leonard}},
\bibnamefont{and}  \bibinfo{author}{\bibfnamefont{C.~A.}~\bibnamefont{Sackett}},
  \bibinfo{journal}{Phys. Rev. A} \textbf{\bibinfo{volume}{83}},
  \bibinfo{pages}{063613} (\bibinfo{year}{2011}).


\bibitem[{\citenamefont{Ullah and Hoogerland}(2011)}]{Ullah2011}
\bibinfo{author}{\bibfnamefont{A.}~\bibnamefont{Ullah}}
  \bibnamefont{and} \bibinfo{author}{\bibfnamefont{M.~D.}~\bibnamefont{Hoogerland}},
  \bibinfo{journal}{Phys. Rev. E} \textbf{\bibinfo{volume}{83}},
  \bibinfo{pages}{046218} (\bibinfo{year}{2011}).

\bibitem[{\citenamefont{Sadgrove and Nakagawa}(2011)}]{Sadgrove2011}
\bibinfo{author}{\bibfnamefont{M.}~\bibnamefont{Sadgrove}}
  \bibnamefont{and} \bibinfo{author}{\bibfnamefont{K.}~\bibnamefont{Nakagawa}},
  \bibinfo{journal}{Rev. Sci. Instrum.} \textbf{\bibinfo{volume}{82}},
  \bibinfo{pages}{113104} (\bibinfo{year}{2011}).

\bibitem[{\citenamefont{Inouye et~al.}(1998)\citenamefont{Inouye, Andrews,
  Stenger, Miesner, Stamper-Kurn, and Ketterle}}]{inouye1998observation}
\bibinfo{author}{\bibfnamefont{S.}~\bibnamefont{Inouye}},
  \bibinfo{author}{\bibfnamefont{M.}~\bibnamefont{Andrews}},
  \bibinfo{author}{\bibfnamefont{J.}~\bibnamefont{Stenger}},
  \bibinfo{author}{\bibfnamefont{H.}~\bibnamefont{Miesner}},
  \bibinfo{author}{\bibfnamefont{D.}~\bibnamefont{Stamper-Kurn}},
  \bibnamefont{and} \bibinfo{author}{\bibfnamefont{W.}~\bibnamefont{Ketterle}},
  \bibinfo{journal}{Nature} \textbf{\bibinfo{volume}{392}},
  \bibinfo{pages}{151} (\bibinfo{year}{1998}).

\bibitem[{\citenamefont{Chin et~al.}(2010)\citenamefont{Chin, Grimm, Julienne,
  and Tiesinga}}]{feshbachrev}
\bibinfo{author}{\bibfnamefont{C.}~\bibnamefont{Chin}},
  \bibinfo{author}{\bibfnamefont{R.}~\bibnamefont{Grimm}},
  \bibinfo{author}{\bibfnamefont{P.}~\bibnamefont{Julienne}}, \bibnamefont{and}
  \bibinfo{author}{\bibfnamefont{E.}~\bibnamefont{Tiesinga}},
  \bibinfo{journal}{Rev. Mod. Phys.} \textbf{\bibinfo{volume}{82}},
  \bibinfo{pages}{1225} (\bibinfo{year}{2010}).

\bibitem[{\citenamefont{Ashcroft and Mermin}(1976)}]{ashcroftss1976}
\bibinfo{author}{\bibfnamefont{N.~W.} \bibnamefont{Ashcroft}} \bibnamefont{and}
  \bibinfo{author}{\bibfnamefont{D.~N.} \bibnamefont{Mermin}},
  \emph{\bibinfo{title}{{Solid State Physics}}} (\bibinfo{publisher}{Brooks
  Cole}, \bibinfo{year}{1976}).

\bibitem[{\citenamefont{Hecht}(2001)}]{hecht2001optics}
\bibinfo{author}{\bibfnamefont{E.}~\bibnamefont{Hecht}},
  \emph{\bibinfo{title}{{Optics}}} (\bibinfo{publisher}{Addison-Wesley},
  \bibinfo{year}{2001}), \bibinfo{edition}{4th} ed.

\bibitem[{\citenamefont{Cataliotti et~al.}(2001)\citenamefont{Cataliotti,
  Scheunemann, H\"ansch, and Weitz}}]{subfourier18}
\bibinfo{author}{\bibfnamefont{F.~S.} \bibnamefont{Cataliotti}},
  \bibinfo{author}{\bibfnamefont{R.}~\bibnamefont{Scheunemann}},
  \bibinfo{author}{\bibfnamefont{T.~W.} \bibnamefont{H\"ansch}},
  \bibnamefont{and} \bibinfo{author}{\bibfnamefont{M.}~\bibnamefont{Weitz}},
  \bibinfo{journal}{Phys. Rev. Lett.} \textbf{\bibinfo{volume}{87}},
  \bibinfo{pages}{113601} (\bibinfo{year}{2001}).

\bibitem[{\citenamefont{Martin et~al.}(2008)\citenamefont{Martin, Georgeot, and
  Shepelyansky}}]{Martin2008}
\bibinfo{author}{\bibfnamefont{J.}~\bibnamefont{Martin}},
  \bibinfo{author}{\bibfnamefont{B.}~\bibnamefont{Georgeot}}, \bibnamefont{and}
  \bibinfo{author}{\bibfnamefont{D.~L.} \bibnamefont{Shepelyansky}},
  \bibinfo{journal}{Phys. Rev. Lett.} \textbf{\bibinfo{volume}{100}},
  \bibinfo{pages}{044106} (\bibinfo{year}{2008}).

\bibitem[{\citenamefont{Fishman et~al.}(2003)\citenamefont{Fishman, Guarneri,
  and Rebuzzini}}]{fishman2003theory}
\bibinfo{author}{\bibfnamefont{S.}~\bibnamefont{Fishman}},
  \bibinfo{author}{\bibfnamefont{I.}~\bibnamefont{Guarneri}}, \bibnamefont{and}
  \bibinfo{author}{\bibfnamefont{L.}~\bibnamefont{Rebuzzini}},
  \bibinfo{journal}{J. Stat. Phys.} \textbf{\bibinfo{volume}{110}},
  \bibinfo{pages}{911} (\bibinfo{year}{2003}).

\bibitem[{\citenamefont{Bach et~al.}(2005)\citenamefont{Bach, Burnett, d'Arcy,
  and Gardiner}}]{bach2005quantum}
\bibinfo{author}{\bibfnamefont{R.}~\bibnamefont{Bach}},
  \bibinfo{author}{\bibfnamefont{K.}~\bibnamefont{Burnett}},
  \bibinfo{author}{\bibfnamefont{M.~B.}~\bibnamefont{d'Arcy}}, \bibnamefont{and}
  \bibinfo{author}{\bibfnamefont{S.~A.}~\bibnamefont{Gardiner}},
  \bibinfo{journal}{Phys. Rev. A} \textbf{\bibinfo{volume}{71}},
  \bibinfo{pages}{033417} (\bibinfo{year}{2005}).

\bibitem[{\citenamefont{Fishman et~al.}(2002)\citenamefont{Fishman, Guarneri,
  and Rebuzzini}}]{fishman2002stable}
\bibinfo{author}{\bibfnamefont{S.}~\bibnamefont{Fishman}},
  \bibinfo{author}{\bibfnamefont{I.}~\bibnamefont{Guarneri}}, \bibnamefont{and}
  \bibinfo{author}{\bibfnamefont{L.}~\bibnamefont{Rebuzzini}},
  \bibinfo{journal}{Phys. Rev. Lett.} \textbf{\bibinfo{volume}{89}},
  \bibinfo{pages}{084101} (\bibinfo{year}{2002}).

\bibitem[{\citenamefont{Andersen and Sleator}(2009)}]{andersen2009lattice}
\bibinfo{author}{\bibfnamefont{M.~F.}~\bibnamefont{Andersen}} \bibnamefont{and}
  \bibinfo{author}{\bibfnamefont{T.}~\bibnamefont{Sleator}},
  \bibinfo{journal}{Phys. Rev. Lett.} \textbf{\bibinfo{volume}{103}},
  \bibinfo{pages}{070402} (\bibinfo{year}{2009}).

\bibitem[{\citenamefont{Magnus}(1954)}]{magnus1954exponential}
\bibinfo{author}{\bibfnamefont{W.}~\bibnamefont{Magnus}},
  \bibinfo{journal}{Commun. Pure Appl. Math.} \textbf{\bibinfo{volume}{7}},
  \bibinfo{pages}{649} (\bibinfo{year}{1954}).

\bibitem[{\citenamefont{Wang}(2005)}]{Wang200517}
\bibinfo{author}{\bibfnamefont{H.}~\bibnamefont{Wang}}, \bibinfo{journal}{Appl.
  Math. Comput.} \textbf{\bibinfo{volume}{170}}, \bibinfo{pages}{17 }
  (\bibinfo{year}{2005}).

\bibitem[{\citenamefont{Sakurai and Tuan}(1994)}]{sakurai1994modern}
\bibinfo{author}{\bibfnamefont{J.}~\bibnamefont{Sakurai}} \bibnamefont{and}
  \bibinfo{author}{\bibfnamefont{S.}~\bibnamefont{Tuan}},
  \emph{\bibinfo{title}{{Modern Quantum Mechanics}}}
  (\bibinfo{publisher}{Addison-Wesley Reading (Mass.)}, \bibinfo{year}{1994}).

\end{thebibliography}
\end{document}